%Paper: hep-ph/9207259
%From: JAFFE@lampf.lanl.gov
%Date: Fri, 24 Jul 1992 17:10:53 -0600 (MDT)

%This is a plain TeX Format.  Figures not included.  Figures are available
%from the author.

\def\lapp{\hbox{$ {     \lower.40ex\hbox{$<$}
                   \atop \raise.20ex\hbox{$\sim$}
                   }     $}  }
\def\rapp{\hbox{$ {     \lower.40ex\hbox{$>$}
                   \atop \raise.20ex\hbox{$\sim$}
                   }     $}  }
\def\thru#1{\mathrel{\mathop{#1\!\!\!/}}}
\def\svec#1{\skew{-2}\vec#1}
\def\ll{\left\langle}
\def\rr{\right\rangle}
\def\pmb#1{\setbox0=\hbox{$#1$}%
\kern-.025em\copy0\kern-\wd0
\kern.05em\copy0\kern-\wd0
\kern-.025em\raise.0433em\box0 }
\font\eightrm=cmr8
\magnification=1200
\hoffset=-.1in
\voffset=-.2in

\vsize=7.5in
\hsize=5.6in
\tolerance 10000

\baselineskip 12pt plus 1pt minus 1pt
\pageno=0
\centerline{{\bf SPIN STRUCTURE FUNCTIONS}\footnote{*}{This
work is supported in part by funds
provided by the U. S. Department of Energy (D.O.E.) under contract
\#DE-AC02-76ER03069.}}
\vskip 24pt
\centerline{R. L. Jaffe}
\vskip 12pt
\centerline{\it Center for Theoretical Physics}
\centerline{\it Laboratory for Nuclear Science}
\centerline{\it and Department of Physics}
\centerline{\it Massachusetts Institute of Technology}
\centerline{\it Cambridge, Massachusetts\ \ 02139\ \ \ U.S.A.}
\vskip 1.5in
\centerline{Invited talk presented at:}
\smallskip
\centerline{\bf Baryon '92}
\smallskip
\centerline{Yale University}
\centerline{New Haven, Connecticut}
\centerline{June 2 -- 4, 1992}
\vfill
\centerline{ Typeset in $\TeX$ by Roger L. Gilson}
\vskip -12pt
\noindent CTP\#2114\hfill July 1992
\eject
\baselineskip 24pt plus 2pt minus 2pt
\noindent{\bf I.\quad INTRODUCTION}
\medskip
\nobreak
The study of spin-dependent effects in deep inelastic scattering originated
with the work of the SLAC--Yale Collaboration headed by Vernon Hughes in the
early 1970's.$^{1,\,2}$  This subject has been reinvigorated by the surprising
results published in 1987 by the European Muon Collaboration$^3$ which also
had a large representation from Yale.  It is highly appropriate, therefore, to
dedicate this talk here at Yale to the fascinating subject of the spin
structure of the nucleon as probed at large momentum transfer in lepton
scattering and related processes.

Deep inelastic processes have long been recognized for their dual roles, on
the one hand as precise quantitative {\it tests of QCD\/} --- largely through
the logarithmic $Q^2$-variation generated by radiative corrections --- and on
the other hand as precise and well-understood probes of hadron structure ---
an approach summarized at the most elementary level by the well-known {\it
parton model\/} interpretation of deep inelastic structure functions.  Often,
when parton model phenomena have clear interpretations, they have led to
surprises and major revisions of our concept of hadrons as bound states of
quarks and gluons.  Three examples come to mind:
\medskip
\item{1.}The ``Momentum Sum Rule,'' recognized in the early 1970's$^4$ and
checked experimentally soon after, showed that about 50\% of the nucleon's
``momentum'' is carried by gluons.  A surprise to quark modelists, this result
is actually {\it predicted\/} by QCD at asymptotic $\ln Q^2$ $(\ln
Q^2/\Lambda^2>\!\!>1)$.$^5$  The fact that it seems to work even at moderate
$Q^2$ suggests either that asymptopia is precocious or that glue is an
intrinsic component of the nucleon bound state.
\medskip
\item{2.}The ``Gottfried Sum Rule,''$^{6}$ which is not a sum rule at all but
instead a measurement of the isospin asymmetry of the antiquarks in the
proton: $\int dx\left( \bar{u}_p (x) - \bar{d}_p(x)\right)$.  The recent NMC
precision measurement of this quantity$^7$ firmly established the existence of
an asymmetry, and requires quark modelists to entertain effects which deform
the sea on account of the valence quarks in a hadron.
\medskip
\item{3.}Finally, and perhaps most significant, the EMC measurement of
$g^{ep}_1(x)$ can be interpreted$^{8,\,9}$ as a direct measure of the fraction
of the nucleon's spin to be found on the spin of the quarks.  The result:
$12.0\pm9.4\pm13.8$\%, caught everyone by surprise.  A reasonably
sophisticated quark model estimate was $60\pm12$\%.$^{8,\,10}$
\medskip
\noindent
Naive quark models are apparently too naive.  Even rather sophisticated models
with relativistic quarks, ambient gluon fields and distant meson clouds do not
easily accommodate these unusual results.

The EMC result, which originates in an
unexpectedly small spin asymmetry at low-$x$ has
stimulated a reconsideration of spin effects in deep inelastic processes. Much
to the surprise of those of us involved in this undertaking, the
reconsideration has led to a much richer and more rational picture of the role
of spin in deep inelastic processes than existed before.  The
interpretation of transverse spin and of chirality in hard
processes\footnote{*}{I use the term {\it hard processes\/} in the same sense
as the fine book of the same title by B.~L.~Ioffe, V.~A.~Khoze and
L.~N.~Lipatov.$^{11}$  The reader should bear in mind that ``hard'' is to be
construed as the opposite of ``soft,'' not ``easy.''  Indeed, ``hard
processes'' are among the few {\it easy\/} aspects of QCD!} has had to be
rewritten.  Along the way, the importance of polarization in deep inelastic
processes with hadronic initial states ({\it e.g.\/} polarized Drell-Yan) has
come to the fore. The principal object of my talk is to outline a unified
description of spin-dependent structure functions which has emerged in the
previous few years.  Much of my work on this subject
was performed in collaboration with
Xiangdong Ji and Aneesh Manohar and my debt to them is substantial.

My talk will be organized as follows:
\medskip
\item{I.}Introduction
\medskip
\item{II.}How to count: enumerating and interpreting the independent structure
functions of hadrons.
\medskip
\item{III.}The $g_1$ saga: some remarks on the longitudinal spin asymmetry
measurement and interpretation.
\medskip
\item{IV.}$h_1(x,Q^2)$: a new structure function and its relation to
chirality, transversity and polarized Drell--Yan.
\medskip
\item{V.}Twist-3: measuring quark-gluon correlations with $g_T(x,Q^2)$ and
$h_L(x,Q^2)$.
\goodbreak
\bigskip
\hangindent=19pt\hangafter=1
\noindent{\bf II.\quad HOW TO COUNT:\hfill\break
 Enumerating and Interpreting the Structure Functions of Hadrons}
\medskip
\nobreak
Merely counting the independent ``structure functions'' of hadrons turns out
to be a non-trivial problem.  Its complexity increases with the degree of
departure from scaling at large-$Q^2$.  Effects which scale (modulo logarithms
of $Q^2$ from radiative corrections) at large-$Q^2$ are ``twist-2.''  Those
which vanish as $1/Q$ (again, mod logarithms) are ``twist-3,'' and so on.  The
discussion here will be {\it complete\/} through twist-3.  Twist-4 effects
which vanish like $1/Q^2$ require a considerable investment in formalism and
are much harder to extract from experiment.

The familiar quark and gluon distribution of the parton model are actually
special cases of {\it general light-cone correlation functions\/}.  A couple
of generic examples are
$$\eqalignno{f(x) &\equiv \int {d\lambda\over 2\pi} e^{-i\lambda x} \ll
P\left| \phi(0) \phi(\lambda n) \right| P\rr &(2.1)\cr
E(x,x') &\equiv \int {d\lambda\,d\lambda'\over 4\pi^2} e^{-i\lambda x -
i\lambda'x'} \ll P\left|\phi(0) \phi(\lambda n) \phi(\lambda'n)\right|P\rr
&(2.2)\cr}$$
where $p^\mu = (p,0,0,p)$, $n^\mu= \left( {1\over 2p},0,0,- {1\over
2p}\right)$, $P^\mu = p^\mu + {M^2\over 2} n^\mu$, $\phi(x)$ is a generic
field, and $p$, which is arbitrary, labels the frame.  $p={M\over 2}$
corresponds to the target rest frame and $p\to\infty$ is the ``infinite
momentum frame.''  Both $f(x)$ and $E(x,x')$ depend non-trivially
on a renormalization scale, $\mu^2$  [For a careful discussion see
Ref.~[12].]  In an asymptotically free theory like QCD in which $\beta(g)\sim
g^3$ for small $g$, the $\mu$-dependence of $f(x)$ and $E(x,x')$ is at most a
power of $\ln\mu$.  These logarithms do not change the structure of our
results and they can be considered separately, hence I will often suppress the
renormalization scale.

Both $f(x)$ and $E(x,x')$ are ground state correlation functions with the
correlation taken along a tangent to the light-cone. $F(x)$ has a simple,
probabilistic interpretation obtained by inserting a complete set of states:
$$f(x) = \sum\limits_X \left| \ll P \left| \phi(0) \right| X\rr\right|^2
\delta\left( P^+_X - (1-x) P^+\right) \eqno(2.3)$$
so $f(x)$ is the {\it probability\/} to find a quantum of $\phi$ with $k^+=
xP^+$ in the target state (see Fig.~1a).  $E(x,x')$ does not have a
probabilistic interpretation (see Fig.~1b).  Some general properties of
distributions such as $E(x,x')$ are developed in Ref.~[13].
\midinsert
\vskip 1.5in
\centerline{(a)\hskip 3in (b)}
\centerline{\it Fig.~1}\endinsert

There are relatively simple rules for enumerating those
quark and gluon distribution
functions which contribute effects at twist-2 ({\it i.e.\/} ${\cal O}(1)$ at
large $Q^2$) and twist-3 ({\it i.e.\/} ${\cal O}(1/Q)$ at large-$Q^2$):$^{14}$
\medskip

\item{1.}{\it Identify independent degrees of freedom when
QCD is canonically quantized on the light-cone.}
\medskip
The solution to this problem is well-known:$^{15}$ The quark field is
decomposed with {\it light-cone projection operators\/}, $P_\pm = {1\over 2}
\gamma^\pm \gamma^\mp$, where $\gamma^\pm \equiv {1\over \sqrt{2}} (\gamma^0
\pm \gamma^3)$ ---
$$\psi_\pm\equiv P_\pm \psi \ \ .\eqno(2.4)$$
The gluon field, $A^\mu$, is decomposed with respect to light cone coordinates
--- $A^\mu \to \left( A^+,A^1,A^2,A^-\right)$.  In the gauge $A^+=0$, then
$\psi_+$ and ${\svec A}_\bot = \left( A^1,A^2\right)$ are independent
variables,
whereas $\psi_-$ and $A_-$ are dependent, constrained variables
$$\eqalign{A_- &= A_- \left[ \psi_+, {\svec A}_\bot\right] \cr
\psi_- &= \psi_- \left[ \psi_+, {\svec A}_\bot\right]\ \ .\cr}\eqno(2.5)$$
The independent degrees of freedom, $\psi_+$ and ${\svec A}_\bot$, carry
helicity $\pm {1\over 2}$ and $\pm 1$, respectively.  The dependent degrees of
freedom $\psi_-$ and $A_-$ carry helicity $\pm {1\over 2}$ and 0,
respectively.
\medskip
\item{2.}{\it To enumerate all twist-2 distributions and exhibit their target
spin dependence, count independent helicity amplitudes for forward
parton-hadron scattering using only independent degrees of freedom.}
\medskip
The generic helicity amplitude shown in Fig.~2 gives a visual depiction of the
generic light-cone correlation function of Eq.~(2.1).  Since there is no
momentum transfer to the target, the amplitude corresponds to forward
scattering and takes place along the $\hat e_3$-axis in coordinate space.
Therefore helicity ($\equiv$ angular momentum about $\hat e_3$) is conserved:
$h+H = h'+H'$ and furthermore $A(h,H\to h',H') = A(-h,-H\to -h',-H')$ and
$A(h,H\to h',H') = A(h',H'\to h,H)$ are consequences of parity and time
reversal invariance, respectively.  With these relations, we enumerate the
independent helicity amplitudes for quarks and gluons scattering from targets
with spin-1/2 and spin-1 in Table~1.
\midinsert
\vskip 2in
\centerline{\it Fig.~2}\endinsert
\midinsert
$$\hbox{\vbox{\offinterlineskip
\def\strut{\hbox{\vrule height 12pt depth 6Pt width 0pt}}
\hrule
\halign{
\strut\vrule#\tabskip 0.1in&
#\hfil &
\vrule#&
${{#}}$\hfil&
\vrule#&
${{#}}$\hfil&
\vrule#\tabskip 0.0in\cr
& \multispan5\hfil$\pmb{A(hH\to h'H')}$\hfil & \cr\noalign{\hrule}
& && \omit\hfil Target Spin-1/2\hfil && \omit\hfil Target Spin-1\hfil &
\cr\noalign{\hrule}
& Quarks && \matrix{\cr
A\bigl(\phantom{-}{1\over 2} & \phantom{-}{1\over 2} &
\to & \phantom{-}{1\over 2} & \phantom{-}{1\over 2}\bigr)\cr\noalign{\vskip
0.1cm}
A\bigl( \phantom{-}{1\over 2} & - {1\over 2} & \to & \phantom{-}{1\over 2} &
-{1\over 2}\bigr) \cr\noalign{\vskip 0.1cm}
A\bigl( -{1\over 2} & \phantom{-}{1\over 2} & \to &\phantom{-}{1\over 2} &
-{1\over 2}\bigr) \cr\cr} &&
\matrix{\cr
A\bigl(\phantom{-}{1\over 2} & \phantom{-}1 & \to & \phantom{-}{1\over 2} &
\phantom{-}1\bigr) \cr\noalign{\vskip 0.1cm}
A\bigl( \phantom{-}{1\over 2} & -1 & \to & \phantom{-}{1\over 2} & -1\bigr)
\cr\noalign{\vskip 0.1cm}
A\bigl(-{1\over 2} & \phantom{-}1 & \to & \phantom{-}{1\over 2} & \phantom{-}0
\bigr) \cr\noalign{\vskip 0.1cm}
A\bigl(\phantom{-}{1\over 2} & \phantom{-}0 & \to & \phantom{-}{1\over 2} &
\phantom{-}0\bigr) \cr\cr} & \cr\noalign{\hrule}
& Gluons && \matrix{\cr A\bigl(\phantom{-}1 & \phantom{-}{1\over 2} & \to &
\phantom{-}1 & \phantom{-}{1\over 2}\bigr) \cr\noalign{\vskip 0.1cm}
A\bigl(\phantom{-}1 & -{1\over 2} & \to & \phantom{-}1 & -{1\over 2}\bigr)
\cr\cr} &&
\matrix{\cr
A\bigl(\phantom{-}1 & \phantom{-}1 & \to & \phantom{-}1 &
\phantom{-}1\bigr)\cr\noalign{\vskip 0.1cm}
A\bigl( \phantom{-}1 & -1 & \to & \phantom{-}1 & -1 \bigr) \cr\noalign{\vskip
0.1cm}
A\bigl(-1 & \phantom{-}1 & \to & \phantom{-}1 & -1\bigr) \cr\noalign{\vskip
0.1cm}
A\bigl( \phantom{-}1 & \phantom{-}0 & \to & \phantom{-}1 & \phantom{-}0\bigr)
\cr\cr} & \cr\noalign{\hrule}}}}$$
\centerline{\it Table~1:\quad Twist-2 helicity amplitudes.}
\endinsert

Each amplitude in Table~1 corresponds to a different quark or gluon
distribution function (a function of $x$ and $\ln q^2$).  There are three
quark and two gluon distributions for a spin-1/2 target:
$$\eqalign{f_1(x) & \propto A\left(\matrix{\phantom{-}{1\over 2} &
\phantom{-}{1\over 2} & \to & \phantom{-}{1\over 2} & \phantom{-}{1\over 2}
\cr}\right) + A \left( \matrix{{1\over 2} & - {1\over 2} & \to &
{1\over 2} & -{1\over 2} \cr}\right) \cr\noalign{\vskip 0.2cm}
g_1(x) & \propto A\left( \matrix{ \phantom{-}{1\over 2} & \phantom{-}{1\over
2} & \to & \phantom{-}{1\over 2} & \phantom{-}{1\over 2} \cr}\right) - A
\left( \matrix{{1\over 2} & -{1\over 2} & \to & {1\over
2} & - {1\over 2} \cr}\right) \cr\noalign{\vskip 0.2cm}
h_1(x) &\propto A\left( \matrix{-{1\over 2} & \phantom{-}{1\over 2} & \to
&\phantom{-}{1\over 2} & -{1\over 2} \cr}\right) \cr}\eqno(2.6)$$

$$\eqalign{G(x) & \propto \phantom{A}
\left( \matrix{ 1 & {1\over 2} & \to & 1 & {1\over
2}\cr}\right) + A \left( \matrix{ 1 & -{1\over 2} & \to & 1 & -{1\over
2}\cr}\right) \cr\noalign{\vskip 0.2cm}
\Delta G(x) & \propto A\left( \matrix{ 1 & {1\over 2} & \to & 1 & {1\over 2}
\cr}\right) - A \left( \matrix{ 1 & -{1\over 2} & \to & 1 & -{1\over
2}\cr}\right) \cr}\eqno(2.7)$$
The spin-average and helicity difference structure functions for quarks
($f_1(x)$ and $g_1(x)$) and gluons ($G(x)$ and $\Delta G(x)$) are well-known.
There is experimental information on $f_1$, $g_1$ and $G$.  The helicity-flip
distribution function $h_1(x)$ is little-known because it does not figure
prominently in electron scattering.  Its interpretation is the subject of
Section~IV below.  Note that there is no helicity-flip gluon distribution at
leading twist.

For a spin-1 target there are several new distribution functions beyond the
obvious generalizations of $f_1$, $g_1$, $h_1$, $G$ and $\Delta G$.  Two are
quadrupole distributions for quarks and gluons,
$$\eqalign{
b_1(x) & \propto A\left( \matrix{{1\over 2} & \phantom{-}1 & \to &
{1\over 2} & \phantom{-}1 \cr}\right) + A\left( \matrix{{1\over 2} & - 1 & \to
& {1\over 2} & -1 \cr}\right) - 2A \left( \matrix{{1\over 2} & \phantom{-}0 &
\to & {1\over 2} & \phantom{-}0 \cr}\right) \cr\noalign{\vskip 0.2cm}
B_1(x) &\propto A\left( \matrix{ 1 & \phantom{-}1 & \to & 1 &
\phantom{-}1\cr}\right) + A \left( \matrix{ 1 & -1 & \to & 1 &
\phantom{-}1\cr}\right) - 2A \left(\matrix{
 1 & \phantom{-}0 & \to & 1 & \phantom{-}0
\cr}\right) \cr} \eqno(2.8)$$
and one is a gluon helicity-flip structure function unique to targets with
$J\ge 1$:
$$\Delta(x) \propto A\left(\matrix{ - 1& 1&\to &1&-1\cr}\right)\ \
.\eqno(2.9)$$
$b_1(x)$ and $\Delta(x)$ have novel properties and interesting implications
for nuclear physics.$^{16}$
\medskip
\item{3.}{\it To enumerate twist-3 distributions which enter inelastic hard
processes, count independent helicity amplitudes for forward parton-hadron
scattering with one-independent and one-dependent degree of
freedom.}\footnote{*}{This rule does not give rise to all twist-3
distributions.  The complete description of twist-3 requires the introduction
of distributions which depend on two variables like Eq.~(2.2).  However, the
only ones which appear to arise at tree level ({\it i.e.\/} ignoring QCD
radiative corrections) in hard processes through order $1/Q$ are the ones
captured by this rule.}
\medskip
The dependent gluon mode is $A^-$ with helicity 0.  The dependent quark mode
is $\psi_-$ with helicity $\pm1/2$.  To distinguish $\psi_-$ from $\psi_+$ in
helicity amplitudes we denote it $\underline{1\over 2}$
in helicity amplitudes.  Table~2
summarizes the independent twist-3 helicity amplitudes for spin-1/2 and spin-1
targets.
\midinsert
$$\hbox{\vbox{\offinterlineskip
\def\strut{\hbox{\vrule height 12pt depth 6Pt width 0pt}}
\hrule
\halign{
\strut\vrule#\tabskip 0.1in&
#\hfil &
\vrule#&
${{#}}$\hfil&
\vrule#&
${{#}}$\hfil&
\vrule#\tabskip 0.0in\cr
& \multispan5\hfil$\pmb{A(hH\to h'H')}$\hfil & \cr\noalign{\hrule}
& && \omit\hfil Target Spin-1/2\hfil && \omit\hfil Target Spin-1\hfil &
\cr\noalign{\hrule}
& Quarks && \matrix{\cr A\bigl(\phantom{-}\underline{1\over 2} &
\phantom{-} {1\over 2} & \to & \phantom{-}{1\over 2} &
\phantom{-}{1\over 2}\bigr) \cr\noalign{\vskip 0.1cm}
A\bigl(\phantom{-}\underline{1\over 2} & - {1\over 2} & \to &
\phantom{-}{1\over 2} & - {1\over 2} \bigr) \cr\noalign{\vskip 0.1cm}
A\bigl( - \underline{1\over 2} & \phantom{-}{1\over 2} & \to &
\phantom{-}{1\over 2} & - {1\over 2}\bigr) \cr\cr} &&
\matrix{\cr A\bigl(\phantom{-}\underline{1\over 2} & \phantom{-}1 & \to &
\phantom{-}{1\over 2} & \phantom{-}1\bigr) \cr\noalign{\vskip 0.1cm}
A\bigl( \phantom{-}\underline{1\over 2} & - 1 & \to & \phantom{-}{1\over 2} &
-1 \bigr) \cr\noalign{\vskip 0.1cm}
A\bigl(-\underline{1\over 2} & \phantom{-}1 & \to & \phantom{-}{1\over 2} &
\phantom{-}0 \bigr) \cr\noalign{\vskip 0.1cm}
A\bigl( -{1\over 2} & \phantom{-}1 & \to & \phantom{-}\underline{1\over 2} &
\phantom{-}0 \bigr) \cr\noalign{\vskip 0.1cm}
A\bigl( \phantom{-}\underline{1\over 2} & \phantom{-}0 & \to &
\phantom{-}{1\over 2} & \phantom{-}0 \bigr) \cr\cr} & \cr\noalign{\hrule}
& Gluons && \matrix{A\bigl(\phantom{-}0 & \phantom{-}{1\over 2} & \to &
\phantom{-}1 & -{1\over 2}\bigr) \cr} &&
\matrix{\cr A\bigl( \phantom{-}0 & \phantom{-}1 & \to &\phantom{-}1 &
\phantom{-}0\bigr) \cr\noalign{\vskip 0.1cm}
A\bigl(\phantom{-}0 & \phantom{-}0 & \to &\phantom{-}1 & -1\bigr) \cr\cr} &
\cr\noalign{\hrule}}}}$$
\centerline{\it Table~2:\quad Twist-3 Helicity Amplitudes}
\endinsert

{}From Table~2 we see that a description of the nucleon requires three twist-3
quark distributions, one spin averaged,
$$e(x) \propto A \left( \matrix{\underline{1\over 2} & {1\over 2} & \to
&{1\over 2} & {1\over 2}\cr}\right) + A\left( \matrix{ \underline{1\over 2} &
-{1\over 2} &\to & {1\over 2} & -{1\over 2} \cr}\right)\ \ ,\eqno(2.10)$$
one helicity difference,
$$h_L(x) \propto A \left( \matrix{\underline{1\over 2} & {1\over 2} &\to &
{1\over 2} & {1\over 2}\cr}\right) - A \left(\matrix{\underline{1\over 2} & -
{1\over 2} & \to & {1\over 2} & -{1\over 2}\cr}\right)\ \ ,\eqno(2.11)$$
and one helicity flip,
$$g_T (x) \propto A \left( \matrix{ -\underline{1\over 2} & {1\over 2} & \to &
{1\over 2} & -{1\over 2}\cr}\right)\ \ .\eqno(2.12)$$
There is only one twist-3 gluon distribution for the nucleon and it involves
helicity flip
$$G_T(x) \propto A\left( \matrix{ 0 & {1\over 2} &\to 1 & -{1\over
2}\cr}\right)\ \ .\eqno(2.13)$$

Some comments on this analysis:
\medskip
\item{$\bullet$}I have suppressed flavor labels. For each twist and helicity
structure there are independent distributions for each flavor of quark and
antiquark.  The flavor singlet quark distributions mix
under $q^2$-evolution with
gluon distributions of the same twist and helicity structure.  Specifically
$f_1$ and $G$, $g_1$ and $\Delta G$, and $g_T$ and $G_T$ mix.  $e$, $h_1$ and
$h_L$ do not mix with gluon distributions.
\medskip
\item{$\bullet$}Distributions associated with target helicity flip are in
actuality measured by orienting the target spin transverse to the scattering
axis-$\hat e_3$.  The transversely polarized state is a linear superposition
of helicity eigenstates ({\it e.g.\/} $|\bot\rangle={1\over \sqrt{2}} \left(
|1/2\rangle + |-1/2\rangle\right)$ and $|\top\rangle = {1\over \sqrt{2}} \left(
|1/2\rangle - |-1/2\rangle\right)$, where $|\bot\rangle$ and $|\top\rangle$ are
transverse spin eigenstates). Then the asymmetry obtained by reversing the
target's transverse spin isolates the helicity flip amplitude.  Thus $h_1$ and
$g_T$ (and $G_T$ and $\Delta$ as well)
are important in processes involving transversely polarized targets.
\medskip
\item{$\bullet$}If we decompose the independent quark field $\psi_+$ into
chirality eigenstates, we find that (of the nucleon distribution functions)
$h_1$, $h_L$ and $e$ are {\it chirally odd\/} ({\it i.e.\/} the quark fields in
the  correlation function of Eq.~(2.1) have {\it opposite chirality\/}) whereas
$f_1$, $q_1$ and $g_T$ are chirally even (the quark fields in Eq.~(2.1) have
the {\it same chirality\/}).
This explains why $h_1$, $h_L$ and $e$ do not mix with
gluon distributions (all of which are chirally even). Further it explains why
$f_1$, $g_1$ and $g_T$ are well-known and much studied whereas $h_1$, $h_L$
and $e$ are poorly-known and neglected: QED and {\it perturbative\/}
 QCD both conserve
chirality (except for small quark mass terms) so $h_1$, $h_L$ and $e$ are
suppressed in deep inelastic electron (muon and neutrino) scattering.  As we
shall see in Section~IV, they are not suppressed in hard processes with
hadronic initial states such as Drell--Yan production of lepton pairs.
\medskip
The results of this section (for a spin-1/2 target like the nucleon) are
summarized in Table~3, where the twist-2 and twist-3 quark and gluon
distributions are  classified according to target spin dependence and
chirality.
\midinsert
$$\hbox{\vbox{\offinterlineskip
\eightrm
\def\strut{\hbox{\vrule height 8pt depth 3.5pt width 0pt}}
\hrule
\halign{
\strut\vrule#\tabskip 0.03in&
\hfil$#$\hfil &
\vrule#&
\hfil#\hfil &
\hfil#\hfil &
\hfil#\hfil &
\vrule#&
\hfil#\hfil &
\hfil#\hfil &
\hfil#\hfil &
\vrule#\tabskip 0.0in\cr
& && \multispan3\hfil \phantom{$^{\big|}$}{\tenrm Twist-2}
${\cal O}(1)$\hfil && \multispan3\hfil{\tenrm Twist-3}
${\cal O}(1/Q)$\hfil & \cr
&  && \multispan3{} && \multispan3{} & \cr
& \matrix{\hbox{\tenrm Spin}\cr\hbox{\tenrm Dependence}\cr} &&
$\matrix{\hbox{Quark}\cr\hbox{Chiral Even}\cr}$ &
$\matrix{\hbox{Quark}\cr\hbox{Chiral Odd}\cr}$ & Gluon &&
$\matrix{\hbox{Quark}\cr\hbox{Chiral Even}\cr}$ &
$\matrix{\hbox{Quark}\cr\hbox{Chiral Odd}\cr}$ & Gluon
&\cr
& && & &  && & & & \cr\noalign{\hrule}
& && & &  && & & & \cr
& \matrix{\hbox{Spin average}\cr
\hbox{(unpolarized target)}\cr} && $f_1(x)$ & --- &
$G(x)$ && --- & $e(x)$ & --- & \cr
& && & &  && & & & \cr\noalign{\hrule}
& && & &  && & & & \cr
& \matrix{\hbox{Helicity asymmetry}\cr
\hbox{(longitudinal polarization)}\cr}
&& $g_1(x)$ & --- & $\Delta G(x)$ && --- &
$h_L(x)$ & --- & \cr
& && & &  && & & & \cr\noalign{\hrule}
& && & &  && & & & \cr
& \matrix{\hbox{Helicity flip}\cr\hbox{(transverse polarization)}\cr}
 && --- & $h_1(x)$ & --- && $g_T(x)$ & --- &
$G_T(x)$ & \cr& && & &  && & & & \cr\noalign{\hrule}}}}$$
\centerline{\it Table~3:\quad Summary of Nucleon Quark and Gluon Distributions
through Twist-3, ${\cal O}(1/Q)$.}
\endinsert
\goodbreak
\bigskip\hangindent=29pt \hangafter=1
\noindent{\bf III.\quad
THE $\pmb{g_1}$ SAGA:\hfill\break Brief Remarks on the Longitudinal Spin
Asymmetry Measurement and \hfill\break Interpretation}
\medskip
The renewed interest in deep inelastic spin phenomena dates from a single
number appearing in a 1987 paper by the European Muon Collaboration.$^3$  In
Ref.~[3] the EMC reported a value for the area under $g_1(x,Q^2)$ for a proton
target at a nominal value of $Q^2=10\,\hbox{GeV}^2$.  In QCD this integral is
related to axial charges measured in $\beta$-decay and to the flavor singlet
axial charge which is otherwise unmeasured:$^{8,\,17}$
$$\int^1_0 dx\, g^{ep}_1(x,Q^2) = {1\over 18} \left\{ (3F+D) \left( 1 -
\alpha_s (Q^2)/\pi\right) + 2 \Sigma(Q^2) \left( 1 + \left( {33-8f\over
32-2f}\right) {\alpha_s(Q^2)\over \pi}\right)\right\} \ \ .\eqno(3.1)$$
$f$ is the number of flavors, $F$ and $D$ are $SU(3)$ invariant matrix
elements of hyperon $\beta$-decay and $\Sigma(Q^2)$ is the renormalization
scale-dependent$^{18}$ flavor singlet axial charge:
$$s^\mu \Sigma(Q^2)= \bigl\langle Ps \bigl|
\sum_{u,d,s}\bar{q} \gamma^\mu \gamma_5 q
\bigr|Ps\bigr\rangle\bigg|_{Q^2} \ \ .\eqno(3.2)$$
$\Sigma(Q^2)$ is also the fraction of the proton's spin to be found on the
spins of its quarks.  Many issues associated with this sum rule and its
interpretation are reviewed in Ref.~[10].

In the most naive, non-relativistic quark model, $\Sigma_{\rm NQM}=1$, because
only the quarks carry spin in the nucleon.  A {\it model independent
analysis\/} assuming only $SU(3)$ symmetry in hyperon $\beta$-decay and {\it
no polarized strange quarks in the nucleon\/} gives$^{8,\,10}$
$\Sigma_{EJ}\cong 0.60\pm0.12$.  The EMC data give
$$\Sigma_{\rm EMC} \left( Q^2_{\rm EMC}\right) = 0.120 \pm 0.094 \pm 0.138\ \
,\eqno(3.3)$$
far from the expectations of any model and arguably compatible with zero.  Many
explanations have been offered by theorists.  Reviews with a variety of biases
can be found in Refs.~[19--21].  Rather than give a (most-likely superfluous)
review here I will restrict myself to a few comments.
\medskip
\item{$\bullet$}Data --- The EMC result relies heavily on the extrapolation to
$x=0$ guided by the four lowest-$x$ points (see Fig.~3).  Confirmation is
urgently needed.  Several new polarized electron and muon experiments are in
the
works:
\medskip
\itemitem{$\rightarrow$} SMC -- The Spin Muon Collaboration now running at
CERN.  They expect to cover roughly the same $x$ and $Q^2$ range as EMC with
better statistics and control of systematic errors.  They will first
measure the longitudinal asymmetry from deuterium$^{22}$ in order to test
Bjorken's sum rule,$^{23}$ so they will not immediately check the EMC result.
They plan to run on polarized hydrogen in the near future.
If one is willing to accept the validity of the Bjorken sum rule (which is an
iron-clad prediction of QCD), then the deuteron data will check the
old result (modulo problems extracting neutron data from a deuteron target).
\medskip
\itemitem{$\rightarrow$} SLAC E142 and E143 --- will study ${}^3$He (E142) and
deuterium and hydrogen (in NH$_3$ and ND$_3$ targets) (E143).  These classic
End Station A experiments will have limited sensitivity to low-$x$ at
large-$Q^2$.  On the other hand, they promise to report data quickly.
\medskip
\itemitem{$\rightarrow$} HERMES --- A large and ambitious program to place
polarized gas targets in the electron beam at HERA, it is awaiting a full
approval, contingent upon demonstrated
polarization of order 50\% in the electron ring.  This facility will be able to
study a variety of targets in various polarization configurations (both
longitudinal and transverse).  They will not have data until the latter half
of this decade.
\medskip
\itemitem{}\quad It seems as though we shall have wait some time for clear
confirmation of the EMC data from electron or muon scattering.  There is,
however, another possible source of information about $\Sigma(Q^2)$:
\medskip
\itemitem{$\rightarrow$} LSND --- A group at LAMPF is presently mounting an
experiment to measure very low energy elastic neutrino scattering from liquid
scintillator.  At very low energies this probes the axial ``charge'' seen by
the $Z^0$-boson --- namely $\bar{u}\gamma^\mu \gamma_5 u-\bar{d}\gamma^\mu
\gamma_5d - \bar{s}\gamma^\mu \gamma_5s$.$^{23}$  (Heavy quarks can be ignored
--- see Ref.~[24].)  Since the first two terms are known from neutron
$\beta$-decay, the experiment measures the polarized strange quark content of
the proton directly.  An earlier version of this experiment using higher
energy neutrinos at BNL --- and therefore subject to further problems of
interpretation --- provided weak confirmation of the EMC result.$^{25}$  LSND
can confirm the EMC data and eventually check the $SU(3)$ symmetry assumptions
which lurk behind Eq.~(3.1).  LSND has its own subtleties, some associated
with the fact that many nucleons in scintillator are bound in carbon nuclei.
An interesting variation suggested by Garvey {\it et al.\/}$^{26}$ may
circumvent some of these difficulties.
\medskip
\item{$\bullet$}The Gluon Controversy --- The triangle anomaly is closely
connected to $\int^1_0 dx\, g^{ep}_1 (x,Q^2)$.  It generates the
$Q^2$-dependence of $\Sigma(Q^2)$.$^{17,\,18}$  In 1988 several groups
suggested$^{27}$ that a proper interpretation of the anomaly alters the sum
rule, Eq.~(3.1).  The claim is that $\Sigma(Q^2)$ measured by experiment
consists of two pieces: one, the ``true'' quark model piece,
$\widetilde{\Sigma}$, is $Q^2$-independent and of order unity; the other,
$-{\alpha_s (Q^2)\over 2\pi}\Delta G(Q^2)$, is a gluonic contribution
proportional to the integrated gluonic spin asymmetry $\Delta G(Q^2)$:
$$\Sigma\left( Q^2\right) = \widetilde{\Sigma} \left( Q^2\right) -
{\alpha_s\left( Q^2\right) \over 2\pi} \Delta G\left( Q^2\right) \ \
.\eqno(3.4)$$
The proposal is summarized in Fig.~4.  The authors of Ref.~[27] suggest that
$\Delta G$ may be so large that it cancels the true quark model piece,
$\widetilde{\Sigma}$, and yields the small result found by EMC.  Carlitz,
Collins and Mueller$^{27}$ suggested ways to measure $\Delta G$ in two jet
production by polarized electrons.
\midinsert
\vskip 2.5in
\centerline{\it Fig.~4}\endinsert

\item{}This
interesting idea generated much activity but has run into a couple of
problems which cloud the original interpretation suggested in Refs.~[27].
First, several groups$^{28}$ pointed out that the separation between
$\widetilde{\Sigma}$ and $\alpha_s \Delta G/2\pi$ is ambiguous.  Second, no
convincing argument has emerged for interpreting $\widetilde{\Sigma}$ as the
true quark model spin contribution.  In fact, a counterexample can be found in
the heavy quark limit$^{10}$ of QCD.  Finally, there is no direct way to
measure $\widetilde{\Sigma}$.  If $\widetilde{\Sigma}$ need not be of order
unity, $\Delta G$ need not be large and the predictive value of Eq.~(3.4)
becomes dubious.  Whatever its significance, Eq.~(3.4) has stimulated
discussion of experiments to measure $\Delta G(x,Q^2)$ which is interesting in
its own right.
\medskip
\item{$\bullet$}$Q\bar{Q}$-Pairs --- There is a prevailing view that the most
natural state for the $Q\bar{Q}$-sea in the nucleon is unpolarized.  A more
careful investigation shows that this is hardly obvious and possibly a
factor in explaining
the small value of $\Sigma_{\rm EMC}$.  In Ref.~[29] Lipkin and I
studied the spin and angular momentum content of a nucleon containing an extra
$Q\bar{Q}$-pair in a variety of constituent quark models.  Our most important
observation was merely that vacuum quantum numbers, $J^{PC} = 0^{++}$, for a
$Q\bar{Q}$-pair require $L = S = 1$.  The $L = S = 0$ state has $J^{PC}=
0^{-+}$ and could only make a nucleon if combined with a less symmetric three
quark state, which in turn would have problems with traditional quark model
successes like magnetic moments and octet axial charges.  Stimulated by this,
we constructed a three-component model of an octet baryon consisting of 1) the
bare three-quark state $|b\rangle$; 2) an additional $0^{++}$ pair
$|b[Q\bar{Q}]^{0^{++}}\rangle$; 3) an additional $1^{++}$ pair
$|b[Q\bar{Q}]^{1^{++}}\rangle$.  The last is the only other $Q\bar{Q}$ state
which can combine with the usual symmetric three-quark state to produce a
nucleon with the proper spin and parity.  We found that it is rather easy to
accommodate the EMC result (and all traditional successes of the quark model)
in such a model, though components 2) and 3) dominate the nucleon's
wavefunction.  Of course this is hardly an {\it explanation\/} of the EMC
data.  It is merely an example of a rather benign sea quark distribution which
accommodates the data.
\medskip
\item{$\bullet$}Flavor tagging to measure $\Delta u$, $\Delta d$ and $\Delta
s$ --- There is great interest in the contributions of individual quark
flavors to the spin asymmetry.  They are defined by
$$\Delta q^a \left( Q^2\right) s^\mu \equiv \bigl\langle
Ps \bigl| \bar{q}^a \gamma_\mu\gamma_5 q^a\bigg|_{Q^2} \big| Ps \bigr\rangle \
\
.\eqno(3.5)$$
The sum rule for $g^{ep}_1$ can be rewritten as a measurement of one linear
combination of $\Delta u$, $\Delta d$ and $\Delta s$.  The separate flavor
contributions cannot be distinguished in electron scattering.  Neutrino
scattering from polarized targets could partially unravel flavor dependence,
but such experiments are out of the questions for the foreseeable future.
\item{}\quad
Recently, Close and Milner have suggested using leading particle effects in
fragmentation to tag the flavor of the struck quark.$^{30}$  The basic
idea is an old one:$^{31}$ a $\pi^+$ observed at large-$x$ and large-$z$ in the
current fragmentation region is most likely a fragment of a
$u$-quark since $\bar{d}$ quarks are suppressed at large $x$.  Similarly
leading $\pi^-$'s are correlated with $d$-quarks and leading $K^\pm$ with
$\bar{s}$ and $s$ (though this correlation is less pronounced since $s$ and
$\bar{s}$ have a soft $x$-distribution). The idea has been studied in
connection with the HERMES proposal$^{32}$ but could be implemented in
any experiment with adequate particle identification.
\goodbreak
\bigskip
\hangindent=28pt\hangafter=1
\noindent{\bf IV.\quad $\pmb{h_1(x,Q^2)}$, CHIRALITY, TRANSVERSITY AND
POLARIZED DRELL--YAN}
\medskip
\nobreak
The twist-2, transverse spin structure function, $h_1(x,q^2)$ has been largely
ignored since its discovery by Ralston and Soper in 1979.$^{33}$ I'd like to
rectify the situation by giving  $h_1(x,Q^2)$ a major place in this
talk.$^{34}$  One
benefit of this will be a clear understanding of transverse spin in the parton
model, a subject which has been surrounded by confusion for years.

$h_1(x,Q^2)$ is a chiral-odd structure function.  It is projected out of a
light-cone quark correlation function with the Dirac matrix
$\sigma_{\mu\nu}\gamma_5$,
$$\eqalign{{1\over 2}\ {1\over 2\pi} \int d\lambda\,e^{i\lambda x} & \ll
Ps\left| \bar{\psi}(0) \sigma_{\mu\nu} i \gamma_5\psi (\lambda n) \right|Ps
\rr\cr
&= h_1(x) \left( s_{\bot\mu} p_\nu - s_{\bot\nu}p_\mu\right)\big/M + h_L (x) M
\left( p_\mu n_\nu - p_\nu n_\mu\right) s\cdot n \cr
&+ h_3 (x) M \left( s_{\bot\mu} n_\nu - s_{\bot\nu} n_\mu\right)
\cr}\eqno(4.1)$$
where $p^\mu = (p,0,0,p)$, $n^\mu = {1\over 2p}(1,0,0,-1)$, $s^\mu \equiv
(s\cdot n) p^\mu + (s\cdot p) n^\mu + s^\mu_\bot$ and $h_L(x)$ and $h_3(x)$
are twist-3 and twist-4 distribution functions, respectively.  If we decompose
$\psi$ into left- and right-handed components, it is clear that $h_1(x)$ is
chirally-odd, as illustrated in Fig.~5a. Deep inelastic lepton scattering in
QCD proceeds via the ``handbag'' diagram, Fig.~5b, and various decorations
which generate $log-Q^2$ dependences, $\alpha_s(Q^2)$ corrections and higher
twist corrections, examples of which are shown in Figs.~5c--f.  All these
involve only {\it chirally-even} quark distributions because the quark
couplings to the photon and gluon preserve chirality.  Only the quark mass
insertion, Fig.~5f, flips chirality.  So up to corrections of order $m_q/Q$,
$h_1(x,Q^2)$ decouples from electron scattering.
\midinsert
\vskip 3.5in
\centerline{\it Fig.~5}\endinsert
\midinsert
\vskip 1.5in
\centerline{\it Fig.~6}\endinsert

There is no analogous suppression of $h_1(x,Q^2)$ in deep inelastic processes
with hadronic initial states such as Drell--Yan. The argument can be read from
the standard parton diagram for Drell--Yan (Fig.~6).  Although chirality is
conserved on each quark line separately, the two quarks' chiralities are
unrelated.  It is not surprising, then, that Ralston and Soper found that
$h_1(x,Q^2)$ determines the transverse-target, transverse-beam asymmetry in
Drell--Yan.

The parton interpretation of $h_1$ can be made transparent by decomposing the
quark fields which appear in Eq.~(4.1) first with respect to the light-cone
projection operator $P_\pm$ and then with respect to various spin projection
operators which commute with $P_\pm$.  The two cases of interest are first, the
chirality projection operators, $P_L$ and $P_R$,
$$P_{{L}\atop{R}} \equiv {1\over 2}\left( 1 \mp \gamma_5\right) \eqno(4.2)$$
which satisfy
$$\left[ P_{{L}\atop{R}}, P_\pm\right] = 0 \ \ ,\eqno(4.3)$$
and second, the transversity projection operators,$^{35}$ $Q_\pm$,
$$Q_\pm \equiv {1\over 2} \left( 1 \mp \gamma_5 \gamma^\bot\right)\ \
,\eqno(4.4)$$
where $\gamma^\bot$ is either $\gamma^1$ or $\gamma^2$, and $Q_\pm$ satisfies
$$\left[ Q_\pm, P_\pm\right] = 0\ \ .\eqno(4.5)$$
As explained in Section~II, $h_1$ involves only independent light-cone degrees
of freedom.  The helicity structure of $h_1$ described in Section II is
apparent in a chiral basis because helicity and chirality coincide up to
irrelevant mass corrections,
$$h_1(x) = {2\over x} \Re \ll P\hat e_1\left| L^\dagger(xP) R(xP)\right|P\hat
e_1\rr \eqno(4.6)$$
compared with $f_1$ and $g_1$
$$\eqalignno{
f_1(x) &= {1\over x} \ll P\left|R^\dagger (xP) R(xP) + L^\dagger (xP) L(xP)
\right|P\rr &(4.7) \cr
g_1(x) &= {1\over x} \ll P\hat e_3\left| R^\dagger (xP) R(xP) - L^\dagger (xP)
L(xP) \right| P\hat e_3\rr \ \ .&(4.8) \cr}$$
In Eqs.~(4.6) -- (4.8) $R(xP)$ and $L(xP)$ are operators which annihilate
independent light cone components of the quark field in eigenstates of $P_R$
and $R_L$, respectively, with momentum $k^+ \equiv xP^+$ and integrated
over ${\svec k}_\bot$.\footnote{*}{The $Q^2$-dependence of $f_1$, $g_1$ and
$h_1$ due to QCD radiative corrections can be restored (to leading logarithmic
order) by integrating over ${\svec k}_\bot$ only up to ${\svec k}^2 \lapp
Q^2$. For a complete discussion, see Ref.~[12].}  According to Eq.~(4.7)
$f_1(x)$ counts quarks with $k^+ = xP^+$ irrespective of helicity, while from
Eq.~(4.8) $g_1(x)$ counts quarks with helicity parallel to the target
helicity minus those antiparallel.  $h_1(x)$ is obscure in this basis.  If,
instead, we diagonalize transversity, then
$$\eqalignno{
f_1 (x) &= {1\over x} \ll P\left| \alpha^\dagger (xP) \alpha(xP) +
\beta^\dagger (xP)
\beta(xP) \right|P\rr &(4.9) \cr
h_1(x) &= {1\over x} \ll P\hat e_1\left| \alpha^\dagger (xP) \alpha(xP) -
\beta^\dagger(xP) \beta(xP) \right| P\hat e_1\rr &(4.10) \cr
g_1(x) &= {2\over x} \Re \ll P\hat e_3\left| \alpha^\dagger
 (xP) \beta(xP) \right|P\hat e_3\rr &(4.11)\cr}$$
where $\alpha(xP)$ and $\beta(xP)$ annihilate independent components of the
quark field in eigenstates of $Q_+$ and $Q_-$, respectively.  From Eq.~(4.10)
it is clear that $h_1(x)$ counts quarks with $k^+ = xP^+$ signed according to
whether their transversity is parallel or antiparallel to the target
transversity.  In this basis the interpretation of $g_1(x)$ is obscure.

{\it The simple structure of Eqs.~(4.6) -- (4.8) and (4.9) -- (4.11) shows
that transverse spin effects and longitudinal spin effects are on a completely
equivalent footing in perturbative QCD.\/} Not knowing about $h_1(x)$, many
authors, beginning with Feynman,$^{36}$ have attempted to interpret
$g_\bot(x)$ as the natural transverse spin distribution function.  Since
$g_\bot(x)$ is twist-3 and interaction dependent, this attempt led to the
erroneous impression that transverse spin effects were inextricably associated
with off-shellness, transverse momentum and/or quark-gluon
interactions.$^{37}$  The resolution contained in the present analysis is
summarized in Table~4 where the symmetry between transverse and longitudinal
spin effects is apparent.  Only ignorance of $h_1$ and $h_L$ prevented the
appreciation of this symmetry at a much earlier date.
\midinsert
$$\hbox{\vbox{\offinterlineskip
\def\strut{\hbox{\vrule height 10pt depth 5pt width 0pt}}
\halign{
#\hfil&
\vrule#&
\hfil$#$\hfil&
#\tabskip .2in&
\hfil$#$\hfil\tabskip 0.0in&
\vrule#\tabskip 0.0in\cr
\strut &\omit& \hbox{Longitudinal} && \hbox{Transverse} &\omit\cr
\strut &\omit& \hbox{Spin} && \hbox{Spin}&\omit\cr
&\omit&\multispan3{\hrulefill}&\omit\cr
\strut Twist-2 && g_1(x,Q^2) && h_1(x,Q^2)&\cr
\strut Twist-3 && h_L(x,Q^2) && g_T(x,Q^2)&\cr
&\omit&\multispan3{\hrulefill}&\omit\cr}}}$$
{\noindent\narrower\narrower\narrower\narrower\it
Table~4:\quad The symmetry of transverse
and longitudinal spin distribution functions.\smallskip}\endinsert

It is useful to summarize some of the known properties of $h_1(x,Q^2)$ and
compare them with analogous properties of $g_1(x,Q^2)$.
\medskip
\item{$\bullet$}Inequalities:
$$\eqalign{\left| g_1(x,Q^2)\right|< f_1(x,Q^2) \cr
\left| h_1(x,Q^2)\right| < f_1(x,Q^2) \cr}\eqno(4.12)$$
for each flavor of quark and antiquark.
\medskip
\item{$\bullet$}Physical interpretation:\quad $h_1(x,Q^2)$ measures
transversity.  It is chirally odd and related to a bilocal generalization of
the tensor operator, $\bar{q}\sigma_{\mu\nu} i \gamma_5 q$.  $g_1(x,Q^2)$
measures helicity.  It is chirally even and related to a bilocal
generalization of the axial charge operator, $\bar{q}\gamma_\mu \gamma_5 q$.
\medskip
\item{$\bullet$}Sum rules:\quad If we define a ``tensor charge''
$$2s^i \delta q^a (Q^2) \equiv \bigl\langle
 Ps\bigl| \bar{q} \sigma^{0i} i \gamma_5
{\lambda^a\over 2} q\bigg|_{Q^2}\bigr| Ps\bigr\rangle\ \ ,\eqno(4.13)$$
where $\lambda^a$ is a flavor matrix and $Q^2$ is a renormalization scale,
then $\delta q^a(Q^2)$ is related to an integral over $h^a_1(x,Q^2)$,
$$\delta q^a (Q^2) = \int^1_0 dx\,\left( h^a_1(x,Q^2) - h^{\bar{a}}_1
(x,Q^2)\right)\eqno(4.14)$$
where $h^a_1$ and $h^{\bar{a}}_1$ receive contributions from quarks and
antiquarks, respectively.  The analogous expressions for $g_1(x,Q^2)$ involve
axial charges,
$$\eqalignno{2s^i \Delta q^a (Q^2)
&\equiv \bigl\langle Ps \bigl| \bar{q} \gamma^i
\gamma_5 {\lambda^a\over 2} q \bigg|_{Q^2} \bigr| Ps\bigr\rangle &(4.15) \cr
\Delta q^a(Q^2) &= \int^1_0 dx\,\left( g^a_1 (x,Q^2) + g^{\bar{a}}_1
(x,Q^2)\right) \ \ .&(4.16) \cr}$$
Note the contrast: $h_1(x,Q^2)$ is not normalized to a piece of the angular
momentum tensor, so $h_1$, unlike $g_1$, cannot be interpreted as the fraction
of the nucleons' spin found on the quarks' spin.  Note the sign of the
antiquark contributions: $\delta q^a$ is charge-conjugation odd, whereas
$\Delta  q^a$ is charge conjugation even.  All tensor charges $\delta q^a$
have non-vanishing anomalous dimensions,$^{38}$ but none mix with
gluonic operators under renormalization.  In contrast, the flavor non-singlet
axial charges, $\Delta q^a$, $a\not=0$, have vanishing anomalous dimension,
whereas the singlet axial charge $\Delta q^0\propto\Sigma$ has an anomalous
dimension arising from the triangle anomaly.$^{17,\,18,\,10}$
\medskip
\item{$\bullet$}Models:\quad $h_1$ and $g_1$ are identical in non-relativistic
quark models, but differ in relativistic models like the bag model --- see
Fig.~7.$^{34}$
\midinsert
\vskip 2in
\centerline{\it Fig.~7}\endinsert

\medskip
\item{$\bullet$}Role in polarized Drell--Yan:\quad $h_1$, $g_1$ and their
twist-3 counterparts $g_T$ and $h_L$ can be measured in lepton-pair production
with appropriately polarized beams and targets (Drell--Yan process).  If both
target and beam are longitudinally polarized,$^{39}$
$$A_{LL} = {\sum\limits_a e^2_a g^a_1 (x) g^{\bar{a}}_1 (y) \over
\sum\limits_a e^2_a f^a_1 (x) f^{\bar{a}}_1(y)}\ \ .\eqno(4.17)$$
If both target and beam are transversely polarized,$^{33}$
$$A_{TT} = {\sin^2 \theta \cos 2\phi\over 1 + \cos^2\theta} \  {\sum\limits_a
e^2_a h^a_1 (x) h^{\bar{a}}_1 (y) \over \sum\limits_a e^2_a f^a_1 (x)
f^{\bar{a}}_1(y)}\eqno(4.18)$$
and if one is longitudinal and the other transverse,$^{34}$
$$A_{LT} =
{2\sin2\theta\cos\phi\over 1 + \cos^2\theta} \  {M\over \sqrt{Q^2}} \
 {\sum\limits_a e^2_a \left( g^a_1 (x) y g^{\bar{a}}_T (y) - x h^a_L(x)
h^{\bar{a}}_1(y)\right) \over
\sum\limits_a e^2_a f^a_1 (x) f^{\bar{a}}_1 (y) }\ \ .\eqno(4.19)$$
In each case, $A_{AB}$ is the appropriate spin asymmetry.  [The
$Q^2$-dependence of the distributions has been suppressed here.]  The
appearance of $h_1(x)$ at leading-twist ({\it i.e.\/} scaling) in Eq.~(4.18)
illustrates its importance in processes in which it is not suppressed by a
chirality selection rule.  The explicit factor of $m/\sqrt{Q^2}$ in Eq.~(4.19)
confirms the twist-3 assignment of both $g_T$ and $h_L$.
\medskip
\item{$\bullet$}Measuring $h_1$ in electroproduction:$^{14}$\quad $h_1(x)$
does not appear at leading twist in electroproduction because the diagrams of
Fig.~5b -- 5e preserve the quark chirality.  The quark mass insertion in
Fig.~5f gives rise to a contribution of order $m/Q$ to the measured
$g_2(x,Q^2)$ obtained from electroproduction off a transversely polarized
target.  $u$ and $d$ quarks are abundant in the nucleon but have very small
(current) masses.  Heavy quarks have non-negligible masses but are rare in the
nucleon.  In all, $h_1(x,Q^2)$ makes a negligible impact on inclusive
electroproduction.  [It is interesting to note, however, that $h_1(x,Q^2)$ was
discovered and its anomalous dimensions calculated in Ref.~[38] on account of
this small contribution to electroproduction.]

\medskip
\item{}
This situation can be changed by observing a particle --- most easily a single
pion --- in the current fragmentation region: $ep\to HX$.  The spin and twist
properties of this process depend on the quark fragmentation function as well
as the distribution function.  If no spin variables are measured in the final
state, then two fragmentation functions can enter: first $\hat f_1(x,Q^2)$
which is the familiar twist-two fragmentation function.  It is chirally even,
and measures the probability for a quark ($a$) to fragment into a
hadron ($H$) with longitudinal momentum fraction $z$.  [I have
suppressed the labels ${a}$ and ${H}$ on $\hat f_1$.]
$\hat f_1(z,Q^2)$ is analogous to the distribution function $f_1(x,Q^2)$.
Continuing the analogy, there is a twist-three, chiral-odd fragmentation
function, $\hat e(z,Q^2)$, analogous to the distribution function $e(x,Q^2)$
described in Section~II.  Polarizing the target transversely to the beam, the
distribution functions $h_1(x,Q^2)$ and $g_T(z,Q^2)$ can enter.  $h_1(x,Q^2)$
can only contribute if $\hat e(z,Q^2)$ provides a compensating chirality flip.
 $g_T(z,Q^2)$ does not flip chirality, and, to leading twist, fragments via
$\hat f_1(z,Q^2)$.  The resulting asymmetry is the sum of two terms ---
$\left[h_1\otimes \hat e\right]\oplus \left[g_T \otimes \hat f_1\right]$
--- each of which is suppressed
by ${\cal O}(1/Q)$ because one twist-three object ($g_T$ or $\hat e$) is
required in each case:
$$A^H_T (x,z,Q^2) \propto {\Lambda\over \sqrt{Q^2}} \ {\sum\limits_a e^2_a
\left[ h^a_1 (x,Q^2) \hat e^{a/H} (z,Q^2) + g^a_T (x,Q^2) \hat f^{a/H}_1
(z,Q^2)\right] \over
\sum\limits_a e^2_a f^a_1 (x,Q^2) \hat f^{a/H}_1 (z,Q^2)}\ \ .\eqno(4.20)$$
Since $\hat f_1(z,Q^2)$ can be measured in $e^+e^-\to HX$ and $g_T (x,Q^2)$
can be measured in {\it inclusive\/}, polarized electroproduction, it is
possible, at least in principle, to extract both $h_1(x,Q^2)$ and $\hat
e(z,Q^2)$ from a measurement of the $x$- and $z$-dependence of the transverse
spin asymmetry in $ep\to eHX$.$^{14}$
\goodbreak
\bigskip\hangindent=23pt\hangafter=1
\noindent{\bf V.\quad TWIST THREE:\hfill\break
 Measuring Quark-Gluon Correlations
with $\pmb{g_T(x, Q^2)}$ and $\pmb{h_L(x, Q^2)}$}
\medskip
\nobreak
The twist-three quark distributions $g_T(x,Q^2)$ and $h_L(x,Q^2)$ are unique
windows into quark gluon correlations in the nucleon.  All higher twist
effects probe quark-gluon correlations.  $g_T$ and $h_L$ are unique in that
they dominate certain observables, in contrast to
generic higher-twist effects which must be
extracted as corrections to QCD fits to leading twist.  For example, consider
electron scattering from a nucleon target polarized at an angle $\alpha$ to
the incident electron beam (Fig.~8).  The spin-dependent part of the
cross-section is given by$^{40}$
$$\eqalign{{d\Delta\sigma\over dx\,dy\,d\phi} = {e^4 \over 4\pi^2 Q^2}
\Biggl\{ & \cos \alpha \left[ \left( 1 - {y\over 2} - {y^2\over 4}
(\kappa-1)\right) g_1 - {y\over 2} (\kappa-1) g_2\right] \cr
-&\sin\alpha\cos\phi \sqrt{ (\kappa-1) \left( 1 - y - {y^2\over 4}
(\kappa-1)\right)} \left( {y\over 2} g_1 + g_2 \right) \Biggr\}
\cr}\eqno(5.1)$$
where $x = Q^2/2P\cdot q$, $y = P\cdot q/ME$, $\kappa = 1 + 4M^2x^2/Q^2$ and
$\phi$ is the dihedral angle between the scattering plane and the plane
defined by the beam and the target spin.  As promised, effects associated with
$g_1$ scale, but effects of $g_T$ ($=g_1+g_2$) fall at least like $1/Q$.  An
experimenter can measure $g_T$ by (first measuring $g_1$, then) orienting his
target at $90^\circ$ to the electron beam.  This should be contrasted with the
elaborate {\it theoretical\/} analysis necessary to isolate higher-twist in
(say) spin averaged electron scattering.  Of course, the experiment is still
non-trivial: the asymmetry is suppressed relative to the rate by ${\cal
O}(1/Q)$ necessitating a high statistics experiment.
\midinsert
\vskip 3in
\centerline{\it Fig.~8}\endinsert

The precision with which the operator product expansion and perturbative QCD
allow us to analyze electron scattering makes $g_T$ and $h_L$ very useful
tools.  Moments of $g_T$ and $h_L$ measure the expectation values of specific,
well-defined local operators.  One must first separate out a
``contamination'' of twist-2 operators from $g_T$ and $h_L$,
namely$^{41,\,34}$
$$\eqalignno{
g_T (x,Q^2) &= \int^1_0 {dy\over y} g_1 (y,Q^2) + \bar{g}_2 (x,Q^2) &(5.2) \cr
h_L (x,Q^2) &= 2x\int^1_0 {dy\over y^2} h_1 (y,Q^2) + \bar{h}_2(x,Q^2) \ \ .
&(5.3) \cr}$$
This leaves $\bar{g}_2$ and $\bar{h}_2$ which depend explicitly on quark
gluon interactions and on the gauge coupling ${g}$, schematically,
$$\eqalignno{
\bar{g}_2(x) &\sim g \ll Ps \left|\bar{q}\gamma_\mu
\widetilde{G}_{\alpha\beta} q
\right| Ps\rr &(5.4)\cr
\bar{h}_2 (x) &\sim g\ll Ps\left| \bar{q} i \sigma^{\alpha\lambda}
\gamma_5 G^\beta{}_\lambda q \right| Ps\rr \ \ .&(5.5) \cr}$$
To be more precise, for example, $\int^1_0 dx \,x^2\bar{g}_1(x,Q^2)$ is
directly related to the nucleon matrix element of the operator,$^{42,\,40}$
$$\theta_{\left[\sigma,\left\{\mu_1\right]\mu_2\right\}}= {g\over 8}
\bar{\psi} \left( \widetilde{G}_{\sigma\mu_1} \gamma_{\mu_2} +
\widetilde{G}_{\sigma\mu_2} \gamma_{\mu_1} \right) \psi\ \ .\eqno(5.6)$$
Similar sum rules for other moments of $\bar{g}_2$ and $\bar{h}_2$ can be found
in Refs.~[42] and [34], respectively.  Admittedly, at the present time we
cannot compute the right-hand side of the sum rules.  However, models and more
ambitious programs like lattice QCD would be aided by experimental information
on matrix elements such as Eq.~(5.6).  A bag model calculation of
$\bar{g}_2(x)$ and $\bar{h}_2(x)$ is shown in Fig.~9 for the sake of a rough
estimate.$^{34}$
\midinsert
\vskip 3in
\centerline{\it Fig.~9}\endinsert

Twist-three distribution functions evolve with $Q^2$ in a more complicated
manner than leading twist ones.  Typically, at leading twist (and leading
order), distributions obey an Altarelli--Parisi equation of the form,
$${d\over d\ln Q^2} f(x,Q^2) = {\alpha_s(Q^2)\over \pi} \int^1_x dy\, P\left(
{x\over y}\right) f(y,Q^2) \eqno(5.7)$$
with a perturbative ``splitting function'' $P(x/y)$.  If $f(x,Q^2)$ is known
at $Q^2 = Q^2_0$ (with $Q^2_0>\!\!>\Lambda^2$) it can be ``evolved'' to another
$Q^2$ using Eq.~(5.7).  At the very least, Eq.~(5.7) allows experimenters to
amalgamate data at a variety of different (large) $Q^2$-values.

The evolution of a twist-3 distribution like $g_T(x,Q^2)$ is more
complicated.$^{42}$  It does not obey a simple evolution equation.  Instead
$g_T(x,Q^2)$ is related to a pair of more general distribution functions,
$G(x,y,Q^2)$ and $\widetilde{G}(x,y,Q^2)$, defined by relations like
$$\eqalign{\int &{d\lambda\over 2\pi}\ {d\mu\over 2\pi} e^{i\lambda x}
e^{i\mu(y-x)}
\bigl\langle Ps\bigl| \bar{\psi}(0) i D^\alpha (\mu n)\thru n \psi (\lambda n)
\bigg|_{Q^2} \bigr| Ps\bigr\rangle \cr
&= 2i\,\epsilon^{\alpha\beta\mu\nu} n_\beta s_\mu p_\nu G (x,y,Q^2) +
\hbox{lower twist}\cr}\eqno(5.8)$$
and similarly for $\widetilde{G}(x,y,Q^2)$.  $G$ and $\widetilde{G}$ are
generic twist-3 parton distributions.$^{34}$  $g_T (x,Q^2)$ is a simple
projection:
$$g_T(x,Q^2) = {1\over 2x} \int^1_0 dy \left[ \widetilde{G}(x,y) + \widetilde
G(y,z) + G(x,y) - G(y,x) \right]\ \ .\eqno(5.9)$$
So twist-3 is inherently much more complicated than $g_T(x,Q^2)$ suggests.  It
is a happy accident that only this particular projection of $G$ and
$\widetilde{G}$ appears in electron scattering.  The difficulty with evolving
$g_T(x,Q^2)$ is that {\it it is $G(x,y,Q^2)$ and $\widetilde{G}(x,y,Q^2)$
which obey Altarelli--Parisi-like integro-differential evolution equations\/}.
 So a measurement of $g_T(x,Q^2)$ at some  $Q^2=Q^2_0>\!\!>\Lambda^2$
does not provide
enough information to predict $g_T(x,Q^2)$ at some other large
$Q^2$.$^{43,\,44}$
This ``impediment to evolution'' was recently stressed in Ref.~[44].

In a recent paper, Ali, Braun and Hiller$^{45}$ have suggested a way around
this problem.  They study the anomalous dimension matrices for all the {\it
local\/} operators which contribute to $G$ and $\widetilde{G}$.  In the
asymptotic limit of $N_c\to\infty$ ($N_c\equiv$ number of colors) and $x\to 1$
they show that $g_T(x,Q^2)$ is an eigenfunction of the matrix evolution
equations.  In simpler terms: for $N_c$ large and $x$ near 1, $g_t(x,Q^2)$
evolves {\it approximately\/} according to an Altarelli--Parisi equation like
(5.1), although the splitting function is not the naive one which would have
been obtained by ignoring the complexity of the problem (as was done, for
example, in Ref.~[46] in the early days of perturbative QCD).  The authors of
Ref.~[45] argue that for $N_c$ not too large and $x$ not too near 1, their
results remain approximately valid.  If they are right then $g_T(x,Q^2)$ can
be evolved with $Q^2$ more or less like a standard distribution function,
making it possible to interpret data taken at a variety of $Q^2$ values in a
systematic way.
\goodbreak
\bigskip
\noindent{\bf VI.\quad CONCLUSIONS}
\medskip
\nobreak
Manipulation of the spin, twist and chirality dependence of deep inelastic
processes offers us a new sensitivity to the details of nucleon structure.
This varied and precise information comes as a consequence of our confidence
in the formalism provided by perturbative QCD --- yet another example of the
adage that yesterday's novelty is today's tool (and tomorrow's background!).
A carefully planned sequence of experiments including deep inelastic
scattering from a variety of targets (proton, deuteron, ${}^3$He, nuclei) in a
variety of spin states, as well as hadronic processes such as polarized
Drell--Yan, can give us much more detailed information on the internal quark
and gluon structure of the nucleon (and nuclei) than we presently possess.
There is no other program which rivals it in precision or clarity of
interpretation within the framework of QCD.  The experiments to date only
scratch the surface of this rich and challenging subject.
\vfill
\eject
\centerline{\bf REFERENCES}
\medskip
\item{1.}M. J. Alguard, {\it et al.\/} {\bf Phys. Rev. Lett.\/} {\bf 37}
(1976) 1261; {\it 41} (1978) 70.
\medskip
\item{2.}G. Braun, {\it et al.\/} {\bf Phys. Rev. Lett.\/} {\bf 51} (1983)
1135.
\medskip
\item{3.}J. Ashman {\it et al.\/}, {\it Phys. Lett.\/} {\bf B206} (1988) 364;
J. Ashman {\it et al.\/}, {\it Nucl. Phys.\/} {\bf B328} (1989) 1.
\medskip
\item{4.}C. H. Llewellyn--Smith, {\it Phys. Rev.\/} {\bf D4} (1971) 2392.
\medskip
\item{5.}See, for example, G. Altarelli, {\it Phys. Rep.\/} {\bf 81C} (1982) 1.
\medskip
\item{6.}K. Gottfried, {\it Phys. Rev. Lett.\/} {\bf 18} (1967) 1174.
\medskip
\item{7.}P. Amaudruz {\it et al.\/}, {\it Phys. Rev. Lett.\/} {\bf 66} (1991)
560.
\medskip
\item{8.}J. Ellis and R. L. Jaffe, {\it Phys. Rev.\/} {\bf D9} (1974) 1444.
\medskip
\item{9.}L. Seghal, {\it Phys. Rev.\/} {\bf D10} (1974) 1663.
\medskip
\item{10.}R. L. Jaffe and A. Manohar, {\it Nucl. Phys.\/} {\bf B337} (1990)
509.
\medskip
\item{11.}B.~L.~Ioffe, V.~A.~Khoze and L.~N.~Lipatov, {\it Hard Processes\/}
(North Holland, Amsterdam, 1985).
\medskip
\item{12.}C. H. Llewelyn--Smith, ``Quark Correlation Functions and Deep
Inelastic Scattering,'' presented at the CAP-NSERC Summer Institute,
Oxford-TP-89/88 (unpublished).
\medskip
\item{13.}R. J. Jaffe, {\it Nucl. Phys.\/} {\bf B229} (1983) 205.
\medskip
\item{14.}R. J. Jaffe and Ji (to be published).
\medskip
\item{15.}J. Kogut and D. E. Soper, {\it Phys. Rev.\/} {\bf D1} (1970) 2901.
\medskip
\item{16.}P. Hoodbhoy, R. L. Jaffe and A. Manohar, {\it Nucl. Phys.\/} {\bf
B312} (1989) 571; R. L. Jaffe and A. Manohar, {\it Nucl. Phys.\/} {\bf B321}
(1989) 343; {\it Phys. Lett.\/} {\bf B223} (1989) 218.
\medskip
\item{17.}J. Kodaira, {\it Nucl. Phys.\/} {\bf B165} (1979) 129; J. Kodaira,
S. Matsuda, T. Muta, T. Uematsu and K. Sasaki, {\it Phys. Rev.\/} {\bf D20}
(1979) 627; J. Kodaira, S. Matusa, K. Sasaki and T. Uematsu, {\it Nucl.
Phys.\/} {\bf B159} (1979) 99.
\medskip
\item{18.}R.~L.~Jaffe, {\it Phys. Lett.\/} {\bf B193} (1987) 101.
\medskip
\item{19.}A. Manohar, ``Lectures at the Lake Louise Winter Institute 1992,''
UCSD preprint UCSD-PTH-92-10.
\medskip
\item{20.}G. G. Ross, in {\it Proceedings of the 1989 International Symposium
on Lepton and Photon Interactions at HIgh Energies\/} (World Scientific,
Singapore, 1990), p.~41.
(1989).
\medskip
\item{21.}A.H. Mueller, ``QCD at high-energies,'' in {\it Proceedings of the
1990 PANIC Conference, Nucl. Phys.\/} {\bf A527} (1991) 137.
\medskip
\item{22.}SMC private communication
\medskip
\item{23.}J. D. Bjorken, {\it Phys. Rev.\/} {\bf D1} (1970) 1376.
\medskip
\item{24.}D. Kaplan and A. Manohar, {\it Nucl. Phys.\/} {\bf B310} (1988) 527.
\medskip
\item{25.}L. A. Ahrens {\it et al.\/}, {\it Phys. Rev.\/} {\bf D35} (1987)
785.
\medskip
\item{26.}G. T. Garvey, S. Krewald, E. Kolbe and K. Langanke, J\"ulich
preprint (1992).
\medskip
\item{27.}A. V. Efremov and O. V. Teryaev, {\it Phys. Lett.\/} {\bf B240}
(1990) 290.; G. Altarelli and G. G. Ross, {\it Phys. Lett.\/} {\bf B212}
(1988) 391; R. D. Carlitz, J. Collins and A. Mueller, {\it Phys. Lett.\/} {\bf
B212} (1988) 229.
\medskip
\item{28.}G. Bodwin and J. Qiu, {\it Phys. Rev.\/} {\bf D41} (1990) 2755; S.
Bass, B. L. Ioffe, N. N. Nikolaev and A. W. Thomas, {\it Journal of the Moscow
Physical Society\/} {\bf 1} (1991) 317, and references therein.
\medskip
\item{29.}R. L. Jaffe and H. Lipkin, {\it Phys. Lett.\/} {\bf B266} (1991)
458.
\medskip
\item{30.}F. E. Close and R. Milner, {\it Phys. Rev. \/} {\bf D44} (1991)
3691.
\medskip
\item{31.}See F. E. Close, {\it Introduction to Quarks and Partons\/}
(Academic Press, Ny, 1978); R. L. Heimann, {\it J. Phys.\/} {\bf G4} (1978)
173.
\medskip
\item{32.}M. Veltri, N. D\"uren, L. Mankiewicz, K. Rith and A. Sch\"afer,
MPI-Heidelberg preprint (1992).
\medskip
\item{33.}J. Ralston and D. E. Soper, {\it Nucl. Phys.\/} {\bf B152} (1979)
109.
\medskip
\item{34.}R. L. Jaffe and X. Ji, {\it Phys. Rev. Lett.\/} {\bf 67} (1991) 552;
{\it Nucl. Phys.\/} {\bf B375} (1992) 527.
\medskip
\item{35.}R. G. Goldstein and M. J. Moravcsik, {\it Ann. Phys.\/} (NY) {\bf 98}
(1976) 128; {\bf 142} (1982) 219; {\bf 195} (1989) 213.
\medskip
\item{36.}R. P. Feynman, {\it Photon-Hadron Interactions\/} (W. A. Benjamin,
Reading, 1972).
\medskip
\item{37.}See, for example, J. D. Jackson, R. G. Roberts and G. G. Ross, {\it
Phys. Lett.\/} {\bf B226} (1989) 159.
\medskip
\item{38.}See the last paper of Ref.~[17], and A. P. Bukhvostov, E. A. Kuraev
and L. N. Lipatov, {\it JETP Lett.\/} {\bf 37} (1983) 482;
{\it Sov. Phys. JETP\/} {\bf 60} (1983) 22, and X. Artru
and M. Mekhfi, {\it Z. Physik\/} {\bf C45} (1990) 669.
\item{39.}F. E. Close and D. Sivers, {\it Phys. Rev. Lett.\/} {\bf 39} (1977)
1116; J. C. Collins and D. E. Soper, {\it Phys. Rev.\/} {\bf D16} (1977) 2219.
\medskip
\item{40.}R. L. Jaffe, {\it Comm. Nucl. Part. Phys.\/} {\bf 14} (1990) 239.
\medskip
\item{41.}W. Wandzura and F. Wilczek, {\it Phys. Lett.\/} {\bf B172} (1975)
195.
\medskip
\item{42.}E. V. Shuryak and A. I. Vainshteyn, {\it Nucl. Phys.\/} {\bf B201}
(1982) 141.
\medskip
\item{43.}A. P. Bukhvostov {\it et al.\/}, Ref.~[38]; P. G. Ratcliffe, {\it
Nucl. Phys.\/} {\bf B264} (1986) 493; I. I. Balitsky and V. M. Braun, {\it
Nucl. Phys.\/} {\bf B311} (1989) 541.
\medskip
\item{44.}X. Ji and C. Chou, {\it Phys. Rev.\/} {\bf D42} (1990) 3637.
\medskip
\item{45.}A. Ali, V. M. Braun and G. Hiller, {\it Phys. Lett.\/} {\bf B266}
(1991) 117.
\medskip
\item{46.}M. A. Ahmed and G. G. Ross, {\it Nucl. Phys.\/} {\bf B111} (1976)
441.
\vfill
\eject
\centerline{\bf FIGURE CAPTIONS}
\medskip
\item{Fig.~1:}Generalized light-cone distribution function: a) the familiar
two-particle case expressed as a probability; b) a generic three-particle
distribution.
\medskip
\item{Fig.~2:}Forward scattering of a parton (quark or gluon) of momentum $k$
and helicity $h$ from a target of momentum $P$ and helicity $H$.
\medskip
\item{Fig.~3:}EMC data on $g_1(x)$ and its integral.
\medskip
\item{Fig.~4:}Graphical representation of gluonic contribution to $\Sigma$.
\medskip
\item{Fig.~5:}Chirality in deep inelastic scattering: a) Chirally odd
contributions to $h_1(x)$; b)--e) Chirally even contributions to deep
inelastic scattering (plus $L\leftrightarrow R$ for
electromagnetic currents); f) Chirality flip by mass insertion.
\medskip
\item{Fig.~6:}Chirality in Drell--Yan (plus $L\leftrightarrow R$) production
of lepton pairs.
\medskip
\item{Fig.~7;}Bag model estimates of $h_1(x)$ and $g_1(x)$.
\medskip
\item{Fig.~8:}Kinematics for polarized deep inelastic lepton scattering from a
spin-1/2 target polarized at an angle $\alpha$ with respect to the beam axis.
\medskip
\item{Fig.~9:}The proton's twist-3 spin-dependent structure functions $g_2$
and $h_2$ in the Bag model: a) $g_2$ and $\bar{g}_2$; b) $h_2$ and
$\bar{h}_2$.

\par
\vfill
\end